# Suppression of $T_c$ in the $(Y_{0.9}Ca_{0.1})Ba_2Cu_{4-x}Fe_xO_8$ system

R. Escamilla[1], T. Akachi

*Instituto de Investigaciones en Materiales, UNAM, 04510 México D.F.*

R. Gómez, V. Marquina, M.L. Marquina, R. Ridaura

*Facultad de Ciencias, UNAM, 04510 México D.F.*

**ABSTRACT**

In this paper, the effects produced by the iron substitutions in the $(Y_{0.9}Ca_{0.1})Ba_2Cu_{4-x}Fe_xO_8$ system on the superconducting and structural properties are studied. The Rietveld-fit of the crystal structure and Mössbauer spectroscopy results of $(Y_{0.9}Ca_{0.1})Ba_2Cu_{4-x}Fe_xO_8$ samples indicate that, the iron atoms occupy the Cu(1) sites of the $(Cu-O)_2$ double chains in fivefold coordination at low iron concentrations. Besides at high iron concentrations the iron atoms occupy the Cu(1) sites of single Cu-O chains and Cu(2) sites in the $CuO_2$ planes of the $(Y_{0.9}Ca_{0.1})Ba_2Cu_4O_8$ phase with structural defects. Simultaneously, as iron concentration increases, a faster decrease of $T_c$ is observed in this material compared with the $YBa_2Cu_{3-x}Fe_xO_{7-\delta}$ system. According to the charge-transfer model proposed for $YBa_2Cu_4O_8$ under pressure, the decrease in the Cu(1)-O(4) bond length in parallel to the increase in the Cu(2)–O(4) bond length may affect the charge transfer mechanism leading to the suppression of $T_c$.

## 1. INTRODUCTION

Among the high-$T_c$ superconductors, the $YBa_2Cu_4O_8$ (Y124, $T_c$ = 80 K) system has been well studied with regard to structural and physical properties. As it is now well-known, the Y124 crystal structure is closely related to that of the $YBa_2Cu_3O_7$ (Y123, $T_c$ = 90 K) with a double chain $(Cu-O)_2$, instead of a single one, along the *b-a*xis of the unit cell. The positions of the Cu atoms in adjacent Cu-O chains differ by a distance b/2 along the *b-a*xis [1,2]; this fact leads to a *c* length parameter longer than in the Y123 by 27.25 Å. The Y substitution by Ca in the Y124 phase gives rise to the increase of the superconducting transition temperature $T_c$ to 90 K for the nominal composition of $(Y_{0.9}Ca_{0.1})Ba_2Cu_4O_8$ [3]. As it was suggested previously [3] the doped $Ca^{2+}$ ions introduce additional holes as charge carriers into the Y124 phase which increase the value of $T_c$.

---

[1] *Corresponding author. rauleg@servidor.unam.mx*

On the other hand, since the discovery of high–temperature superconductors many experiments have been carried out in order to determine the effects of Cu substitution by other 3d metals have on the superconductivity. It was hoped that these investigations would reveal clues on the mechanism causing the high transition temperatures in these materials. However, it is still not clear what causes the suppression of $T_c$ as dopant concentration is increased. Among the most common effects proposed in the literature are magnetic pair breaking, change in the local symmetry and purely electronic dynamical mechanisms [4]. In any case, basic issues such as dopant site localization and charge state must be considered. Mössbauer spectroscopy has been proven a powerful tool in the determination of the oxidation state of the substituent Fe atoms as well as their environment (i.e., crystallographic site in the structure). Although it is now apparent that the details of the sample as well as on oxygen stoichiometry, a considerable amount of experimental evidence appears to be at least qualitatively well established.

In the present research we have carried out systematic substitutional studies on $(Y_{0.9}Ca_{0.1})Ba_2Cu_{4-x}Fe_xO_8$ (x = 0.025, 0.05, 0.075, 0.1 and 0.2) to clarify where the iron atoms are situated in the structure how they affect its crystalline structure and transport properties. We present X-ray diffractograms, Mössbauer spectroscopy, resistance *vs.* temperature and AC susceptibility *vs.* temperature measurements results. In order to attain a site assignment, the Mössbauer results were complemented with a Rietveld refinement of the X-ray diffraction data.

**2. EXPERIMENTAL**

We have synthesized $(Y_{0.9}Ca_{0.1})Ba_2Cu_{4-x}Fe_xO_8$ (YCa124Fe) samples with x = 0, 0.025, 0.05, 0.075, 0.1 and 0.2 with ambient oxygen pressure using a combined method described as follows: a) powders of $Y_2O_3$, $CaCO_3$, $BaCO_3$, CuO and $^{57}Fe_2O_3$ (99.999%) were mixed in stoichiometric amounts, diluted in 10 ml of nitric acid (96 %) and heated until all the liquid evaporated; the resulting green paste was fired at 700 °C in air to obtain small grain (1-10 μm) precursors. b) The calcinated powders were mixed with sodium nitrate powder (20 % in weight) as catalyst to accelerate the reaction, pressed into pellets, heat treated at 810 °C during 20 h in oxygen flow and slowly cooled (0.5 °C/min) to room temperature. The heat treatment was repeated at least four times.

Phase identification of the samples was done with an X-ray diffractometer Siemens D5000 using Cu-K$_\alpha$ radiation and a Ni filter. Intensities were measured in steps of 0.02° for 14 seconds in the 2θ range 20° – 70° at room temperature and crystallographic phases were identified by comparison with the X-ray patterns of the JCPDS database. Crystallographic parameters were refined using a Rietveld refinement program, RIETQUAN v2.3 [5], with multi-phase capability. The superconducting transition temperatures were determined in a closed-cycle helium refrigerator by measuring resistance *vs.* temperature curves by the standard four-probe technique. The AC susceptibility magnetic was measured upon warming from a zero-field-cooled (ZFC) between 12 and 300 K with a fixed frequency of 125 Hz and amplitude of 1.1 mT (APD Cryogenics Inc. Superconductor Characterization Cryostat). The room temperature Mössbauer spectra were recorded in transmission geometry with a constant acceleration spectrometer, using a $^{57}$Co in Rh source. The spectra were fitted with a constrained-least-squares program.

## 3. RESULTS AND DISCUSSION

Figure 1 shows X-ray diffraction patterns for the $(Y_{0.9}Ca_{0.1})Ba_2Cu_{4-x}Fe_xO_8$ (YCa124Fe) samples synthesized. The analysis of these data indicate that all samples correspond to the Y124 structure although for $0 \leq x \leq 0.075$ (low concentration) faint features of CuO (IDDE nº 05-661) appear and for $x \geq 0.1$ (high concentration) additional peaks corresponding to $(Y_{0.9}Ca_{0.1})Ba_2Cu_3O_y$ (YCa123) and to $BaCuO_2$ (IDDE nº 38-1402) impurities can be also observed. Yanagizawa *et al.* [6] have shown, by High Resolution Transmission Electron Microscopy (HRTEM), that at around a 3 % level of iron doping, a destabilization of the (Cu-O)$_2$ double chains of Y124 takes place, which produces random distributed Cu-O single chains in the Y124 structure. These structural defects are manifest in the X-ray diffraction pattern for the presence of traces associated to "Y123" pseudo-phase. This effect would be difficult to understand if the site occupancy of the iron atoms is other one than the Cu(1) sites. In this context, the Y124 phase with structural defects is a "Y123" pseudo-phase.

Figure 2 shows the normalized resistance *vs.* temperature of the studied samples. We observed that T$_c$ decreases continuously increasing the iron concentration (x), The zero resistance of the sample with x = 0.2, is an extrapolated value, and it is indicated as a solid square in the inset of Fig. 2. In this inset, the $T_c$ variation with x for this system compared with the YBa2Cu4-xFexO8 (Y124Fe) [7] and $YBa_2Cu_{3-x}Fe_xO_y$ (Y123Fe) [8,9] systems. We observed that the $T_c$ values in YCa124Fe system are larger than those reported in the reference [7]. According to Miyatake *et al.* [3] the reason of this increase in $T_c$ is the introduction of holes as charge carriers due to the substitution of Ca ions in the Y sites. On the other hand, we also observed in the Y124Fe and YCa124Fe systems, that at x ≤ 0.1, the rate of decrease of $T_c$ with iron concentration is linear. But at x = 0.2, the YCa124Fe system continues being superconductor ($T_c$ ~ 6 K) while in the Y124Fe system the $T_c$ is suppressed. Fig. 3 shows the magnetization as a function of temperature for the YCa124Fe samples. We observed that the values of the transition temperature determined from the magnetization measurement are in good agreement with the transition temperatures from the resistivity *vs.* temperature curves.

The diffraction patterns of YCa124Fe samples were Rietveld–fitted using an orthorhombic structure with space group *Ammm* (n$^o$ 64). In the refinement process: *a)* It was assumed that Ca atoms substitute the Y sites, *b)* the presence of different impurities, such us $BaCuO_2$, CuO phases and "YCa123" pseudo-phase were considered, *c)* the occupancy factor of Cu and Fe atoms in the Cu(1) and Cu(2) sites were varied independently in order to allow the program to establish the site occupancy of the Fe atoms in the YCa124 and "YCa123" structures, *d)* we permitted the possibility of extra oxygen atoms in the structure: O(5), between two adjacent double chains planes in front of a Cu(1) site, along the *a*-axis. *e)* finally, the isotropic thermal parameters (in Å$^2$) were kept fixed at the values obtained by Adachi *et al* [10]. In Figure 4a, the results for the sample with x = 0.025, as representative of the low concentration samples are shown. In similar way, Figure 4b presents the results for sample with x = 0.2 as representative of the high concentration samples. In Table I the structural parameters and occupation factors obtained for x = 0.025 and x = 0.2 are summarized. O(1) refers to the oxygen atoms along the double chains, O(2) and O(3) to the oxygen atoms in the $CuO_2$ planes and O(4) to the oxygen atoms in the Ba atoms plane. Our results of Rietveld–fit, show the *a*-axis and *b*-axis lattice parameters increases slightly, as a function of increasing (x) and the *c*-axis one decreases continuously (see Figure 5). We observed that the structure of the YCa124Fe samples remains orthorhombic up to the highest Fe doping (x=0.2).

On the other hand, the refinement of atomic positions and occupation factor shows that at low iron concentration the iron atoms are fully localized on the Cu(1) site of the double chains of the YCa124Fe structure. All attempts to put some iron on the Cu(2) site lead to a negative value of this parameter after refinement. At high iron concentrations, our results shown that the iron atoms substitute the Cu(1) sites of the YCa124Fe structure and the Cu(1) and Cu(2) sites of the "YCa123Fe" (see Table I). These replacements originate significant changes in the Cu-O bond lengths. Figure 6 show the Cu(1) – O(4) and Cu(2) – O(4) bond lengths obtained from the refined atomic positions. We observed the former decreases whilst the latter increases with iron concentration. Moreover, with increasing iron concentration, the occupation number for O(5) oxygen of the YCa124Fe structure is increased in good agreement with previous studies in the $YBa_2Cu_{3-x}Co_xO_{7-y}$ system [11]. In this case, for low cobalt concentration, the Co atoms preferentially occupy the Cu(1) site and simultaneously attracting extra oxygen into these layers in an amount consists with $Co^{3+}$. As consequence of this replacement, the Cu(1) – O(4) and Cu(2) – O(4) bond lengths change in the same way.

To supplement the site assignment of the iron atoms in the YCa124Fe structure, we have performed studies of Mössbauer spectroscopy. Figure 7a and Figure 7b shown the Mössbauer spectra for x = 0.025 and x = 0.2 respectively. Each of the spectra for the three lower iron concentrations consist of a single quadrupole doublet -labeled A in Figure 7a -whose parameters (quadrupole splitting $\Delta Q$, isomer shift (IS) and line width $\Gamma$) remain essentially constant (see Table II). However, the spectra of the two higher iron concentration samples consist of four quadrupole doublets -labeled A, B, C and D (see Figure 7b), the most prominent of which (A) corresponds to the single doublet observed at low iron concentrations. Our study show that at low and high iron concentrations, the Mössbauer spectra observed have $\Delta Q$´s values close to those observed in the Y124Fe system (see Table II) [7]. These results suggest that the $\Delta Q$´s values are independent of the substitution of Ca atoms (10%) in the Y sites. Based on the above mentioned, we propose the following assignment: at low iron concentrations, the 0.79 mm/s quadrupole doublet (A) value is associated with $Fe^{3+}$ atoms in the Cu(1) sites in fivefold coordination, resulting from extra oxygen atoms, named O(5), that are attracted by Fe atoms at the Cu(1) sites, and occupy sites placed along the a-axis between two (Cu-O)$_2$ double chains (see Table I), similar site assignment of the iron atoms has been shows by Felner *et al.* [12,13].

As it is now well known, the Mössbauer spectra reported in the literature for the Y123Fe phase consist of three main quadrupole doublets, which have $\Delta Q$'s of approximately 2.0 mm/s, 1.10 mm/s and 0.55 mm/s. The only doublet that has been associated with the Cu(2) sites is the one whose $\Delta Q$ is around 0.55 mm/s [14]. At high iron concentrations, besides of the presence of quadrupole doublet (A), the quadrupole splittings of doublets B (2.0 mm/s) and C (0.90 mm/s) have values close to those associated with $Fe^{3+}$ occupying Cu(1) of the Cu-O single chains of the Y123 structure that result from the destabilization of the (Cu-O)$_2$ double chains [9]. The doublet (B) is associated with $Fe^{3+}$ in the Cu(1) sites with fourfold (planar) coordination and the doublet (C), associated with $Fe^{3+}$ also in the Cu(1) sites with fourfold (non planar) coordination and/or with fivefold (pyramidal) coordination. While the doublet (D) with $\Delta Q$ of around 0.56 mm/s is associated with the Cu(2) sites.

In order to find a more in-depth correlation between the structure and $T_c$ of the YCa124Fe system, the evolution of $T_c$ and Cu(1) – O(4) bond length as a function of the iron concentration (x) is displayed in Figure 8. We observed that the Cu(1)–O(4) bond length decreases as the iron concentration is increased. The change in bond lengths and $T_c$ with the iron concentration, suggests that significant charge-transfer effects occur. It has been claimed that the enhancement of $T_c$ in Y124 under pressure is due mainly to the shortening of the Cu(2) – O(4) bond length and relative elongation of the Cu(1) – O(4) bond length. These changes in the bond lengths were related to the charge-carriers transfer (using the bond valence sum formalism [15,16]) from the Cu-O chains to the CuO$_2$ plane, which causes an increase in charge carriers in the conducting plane CuO$_2$ leading to the increase in $T_c$ [17-19].

Comparing ours results with Y124 system under pressure, we observed that the substitution of Fe atoms in Cu(1) sites, induces an increase in the Cu(2) – O(4) bond length and a decrease in the Cu(1) – O(4) bond length. Simultaneously the $T_c$ decreases with increase iron concentration. According to the charge-transfer model proposed for Y124 under pressure, this behavior in the Cu(1) – O(4) and Cu(2) – O(4) bond lengths of YCa124Fe system may induce an opposite charge transfer, i.e. charge transfer form the CuO$_2$ plane to the double chains (Cu-O)$_2$. This opposite charge transfer can reduce the charge carriers in the conducting CuO$_2$ plane leading to the suppression of $T_c$. This may be one of the reasons for the $T_c$ decrease in YCa124Fe.

## 4. CONCLUSIONS

We have presented a detailed crystallographic, transport and magnetic characterization of the $(Y_{0.9}Ca_{0.1})Ba_2Cu_{4-x}Fe_xO_8$ (YCa124Fe) system by means of XRD powder at room temperature, Mössbauer spectroscopy, resistance *vs.* temperature curves and magnetization measurements. Our results of Mössbauer spectroscopy and Rietveld-fit, indicate that at low iron concentrations the iron atoms occupy partially the Cu(1) sites of the double (Cu-O)$_2$ chains and at high iron concentrations also occupy partially the Cu(1) and Cu(2) sites of the YCa124Fe with structural defects ("YCa123Fe"). These replacements induce significant changes both in the Cu(1)-O(4) and Cu(2)–O(4) bond lengths, the former decreases whilst the latter increases with iron concentration. Simultaneously our resistance *vs* temperature and AC susceptibility magnetic measurements show that the transition temperatures of the YCa124Fe samples decrease with the increasing iron concentration x. The change in bond lengths and $T_c$ with the iron concentration, suggest that significant charge-transfer effects are occurring. According to the charge-transfer model proposed for Y124 under pressure, one should expect a less effective charge transfer in the YCa124Fe system, leading to the suppression of $T_c$. In summary, these facts suggest that the substitution of Fe atoms in the Cu(1) sites modifies the charge transfer mechanism through changes in the Cu(2)–O(4) and Cu(1)–O(4) bond lengths.

**ACKNOWLEDGMENTS**

We would like to thank Dr. Jose Manuel Gallardo-Amores and Dr. Francisco Morales Leal for helping in the manuscript preparation.

**FIGURE CAPTIONS**

Figure 1. X-ray diffraction patterns of the $(Y_{0.9}Ca_{0.1})Ba_2Cu_{4-x}Fe_xO_8$ (YCa124Fe) samples. The symbols represent: (*) CuO, (+) $BaCuO_2$, (o) "YCa123Fe"

Figure 2 Normalized resistance *vs*. temperature of $(Y_{0.9}Ca_{0.1})Ba_2Cu_{4-x}Fe_xO_8$ (YCa124Fe) samples. The inset shows the $T_c$ variation with iron concentration x.

Figure 3 Magnetization as a function of temperature for the $(Y_{0.9}Ca_{0.1})Ba_2Cu_{4-x}Fe_xO_8$ (YCa124Fe) samples

Figure 4 a) Rietveld refinement on the X-ray diffraction pattern for the x = 0.025 sample. b) Rietveld refinement on the X-ray diffraction pattern for the x = 0.2 sample. Experimental pattern(dots), calculated pattern (continuous line), their difference (middle line) and the calculated peaks positions (bottom).

Figure 5 Crystal lattice parameters as a function of iron concentration x.

Figure 6 Cu(1)-O(4) and Cu(2)-O(4) bond lengths *vs*. iron concentration x.

Figure 7 Mössbauer spectra a) for x = 0.025 and b) x = 0.2

Figure 8 $T_c$ *vs*. Cu(1)-O(4) bond length for each concentration of iron.

**TABLE CAPTIONS**

Table I Rietveld refinement results for the $(Y_{0.9}Ca_{0.1})Ba_2Cu_{4-x}Fe_xO_8$ (YCa124Fe) structure with a) x = 0.025 and b) x = 0.2

Table II Mössbauer parameters of the $(Y_{0.9}Ca_{0.1})Ba_2Cu_{4-x}Fe_xO_8$ (YCa124Fe) samples

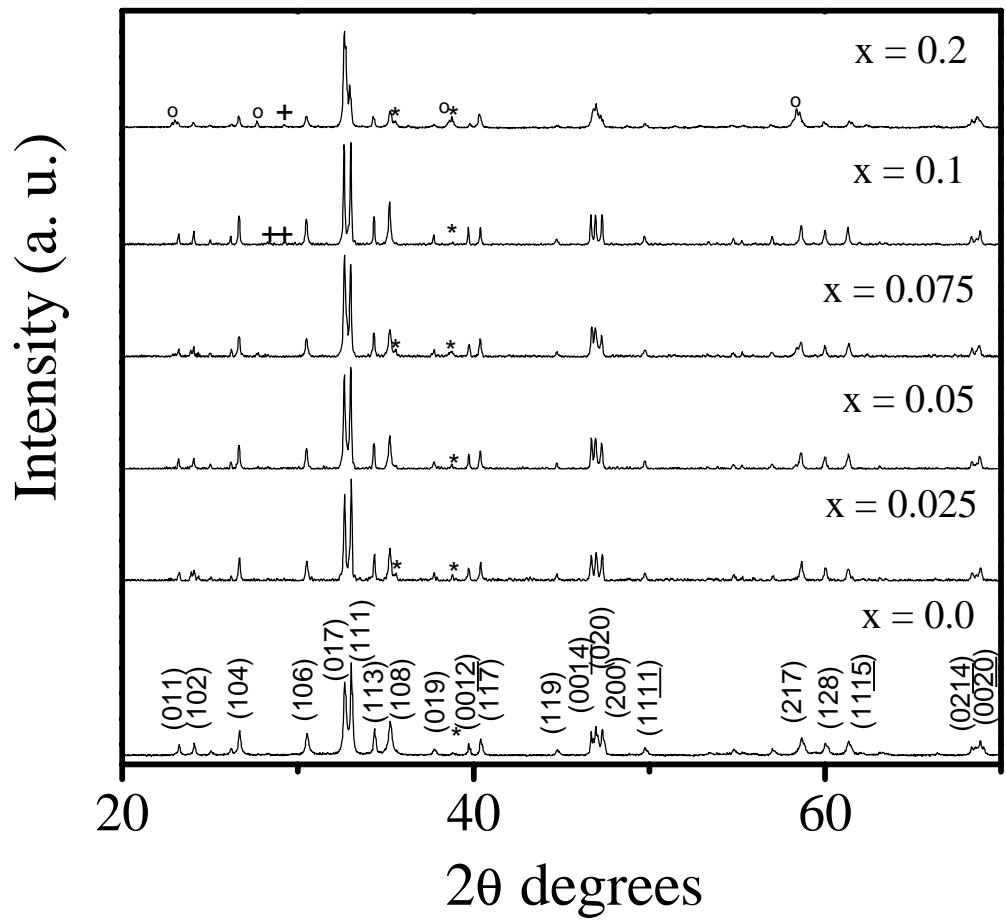

**Figure 1**

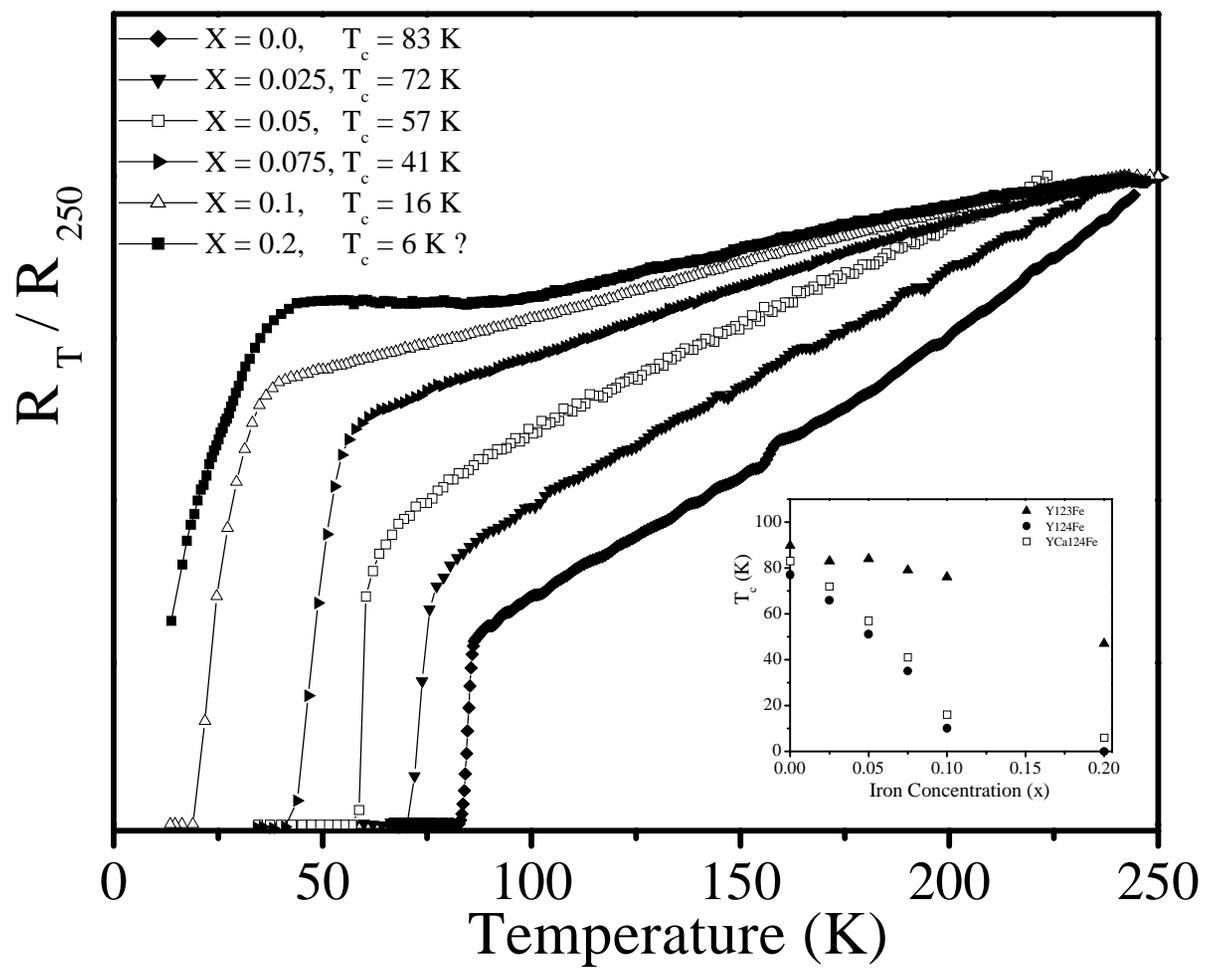

**Figure 2**

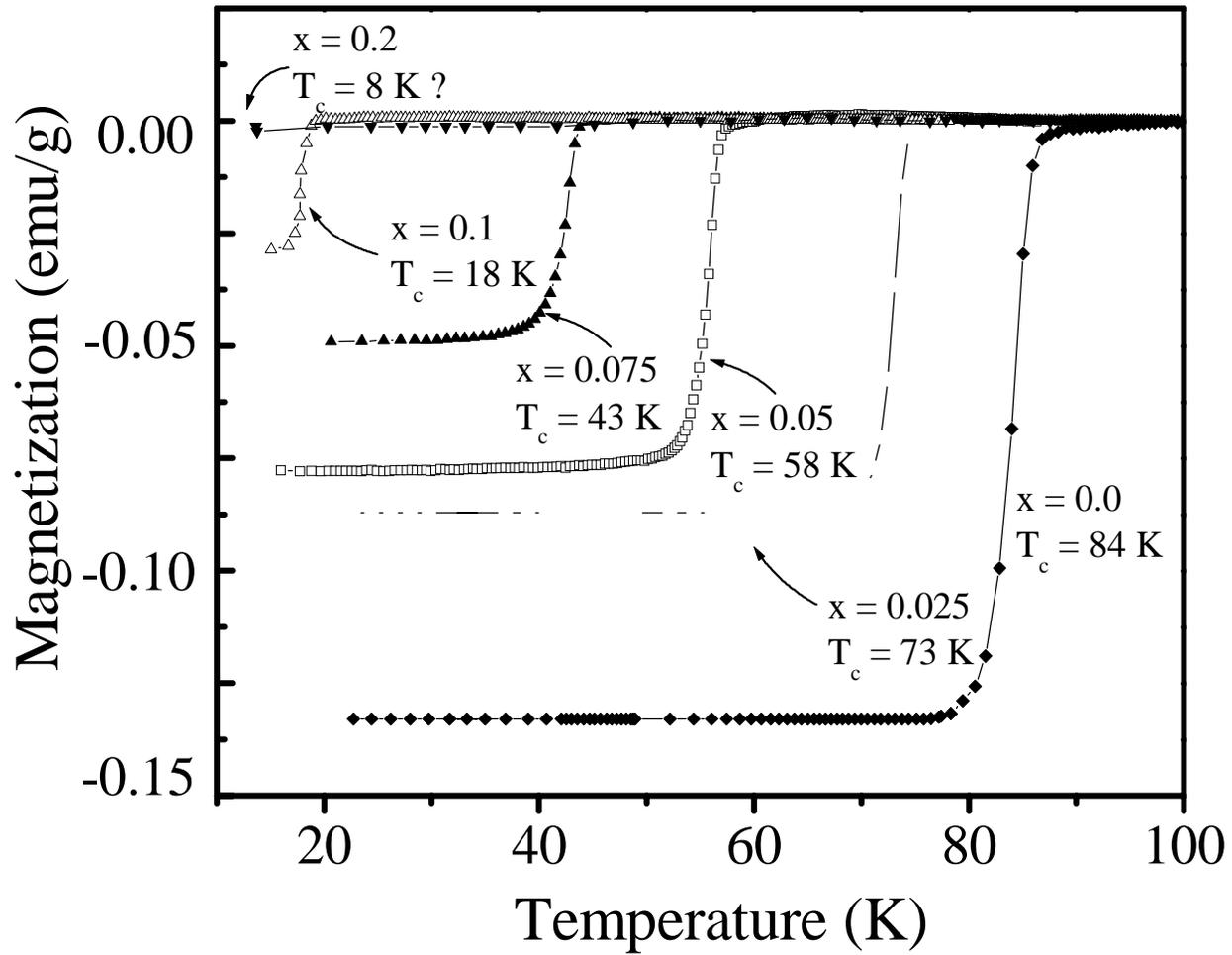

**Figure 3**

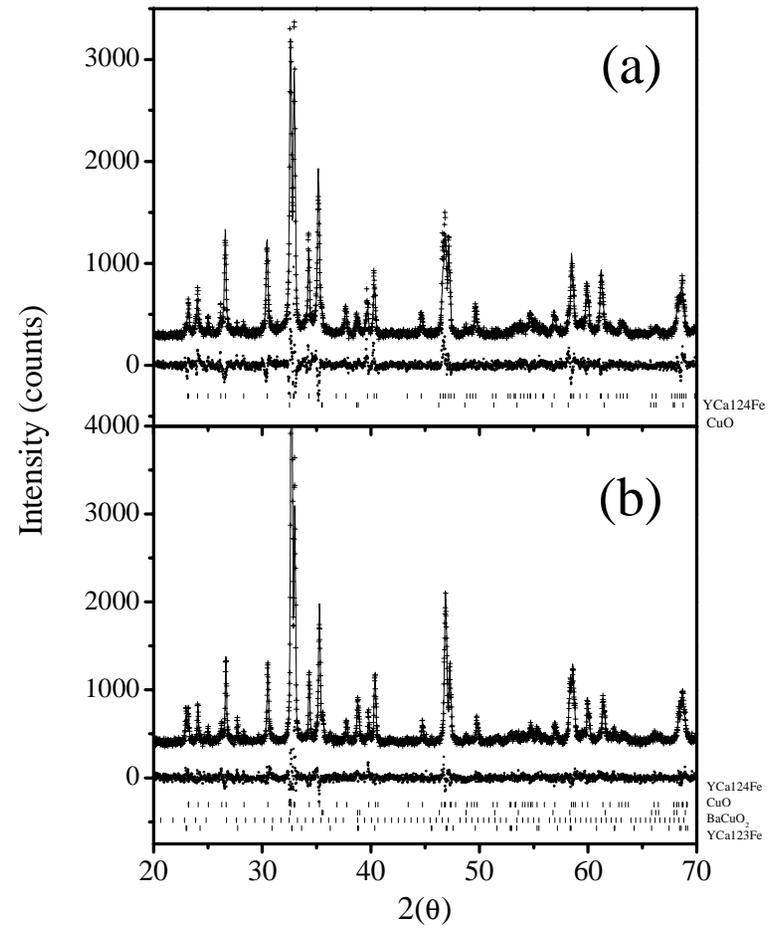

**Figure 4**

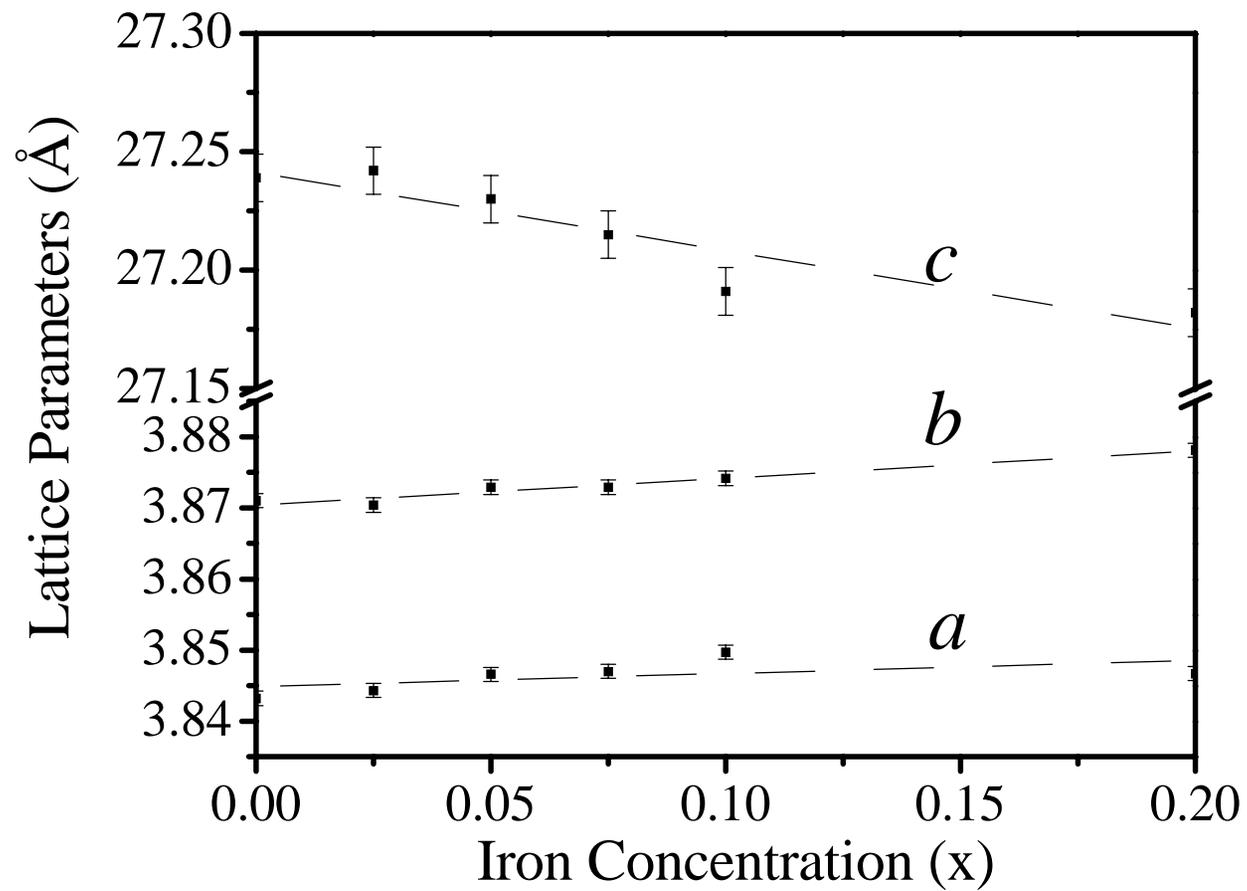

**Figure 5**

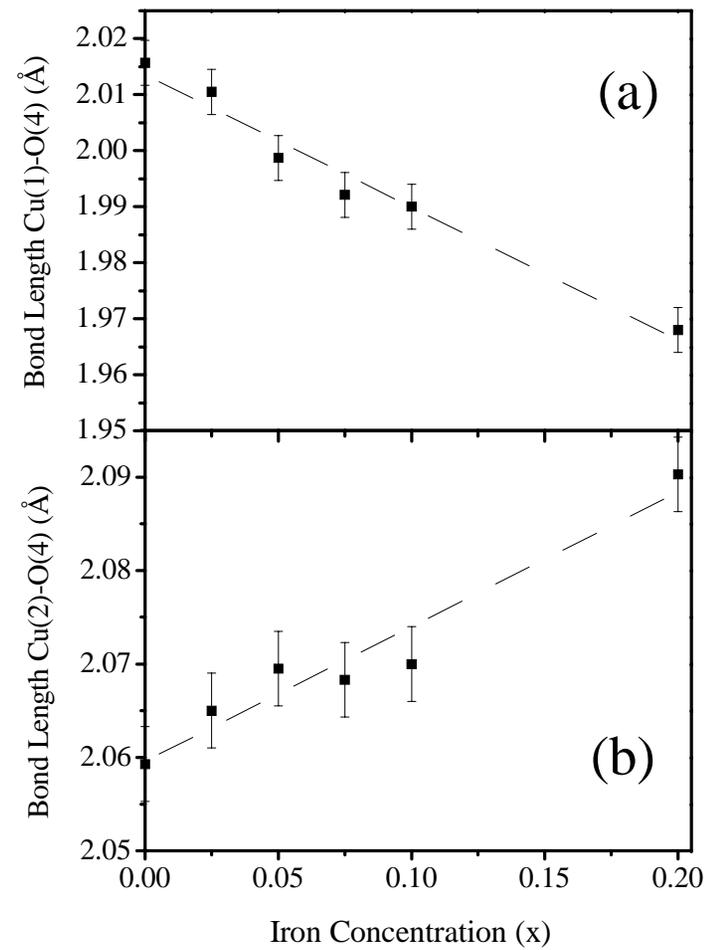

**Figure 6**

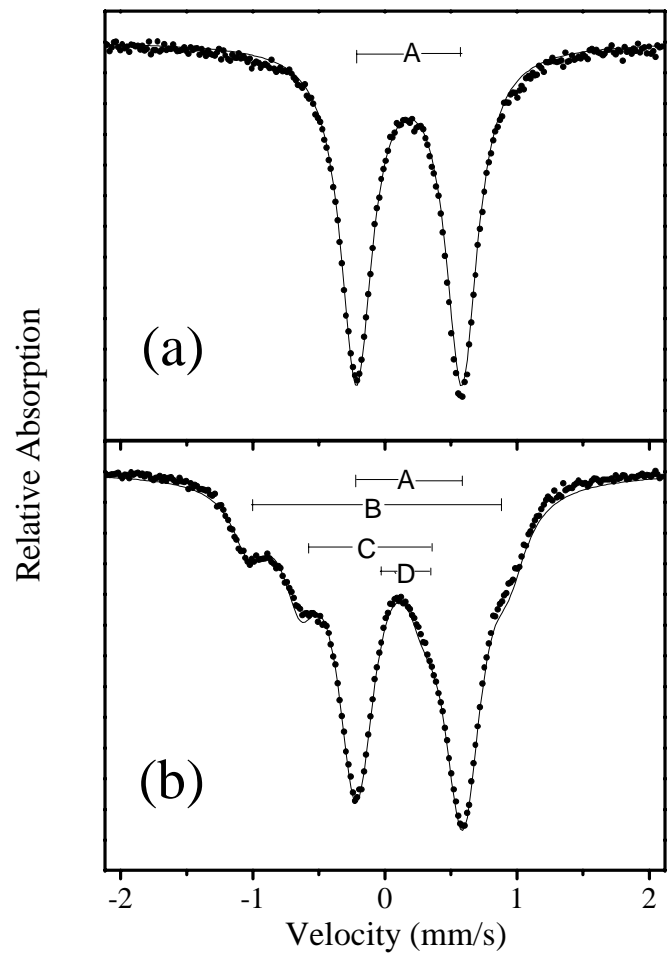

**Figure 7**

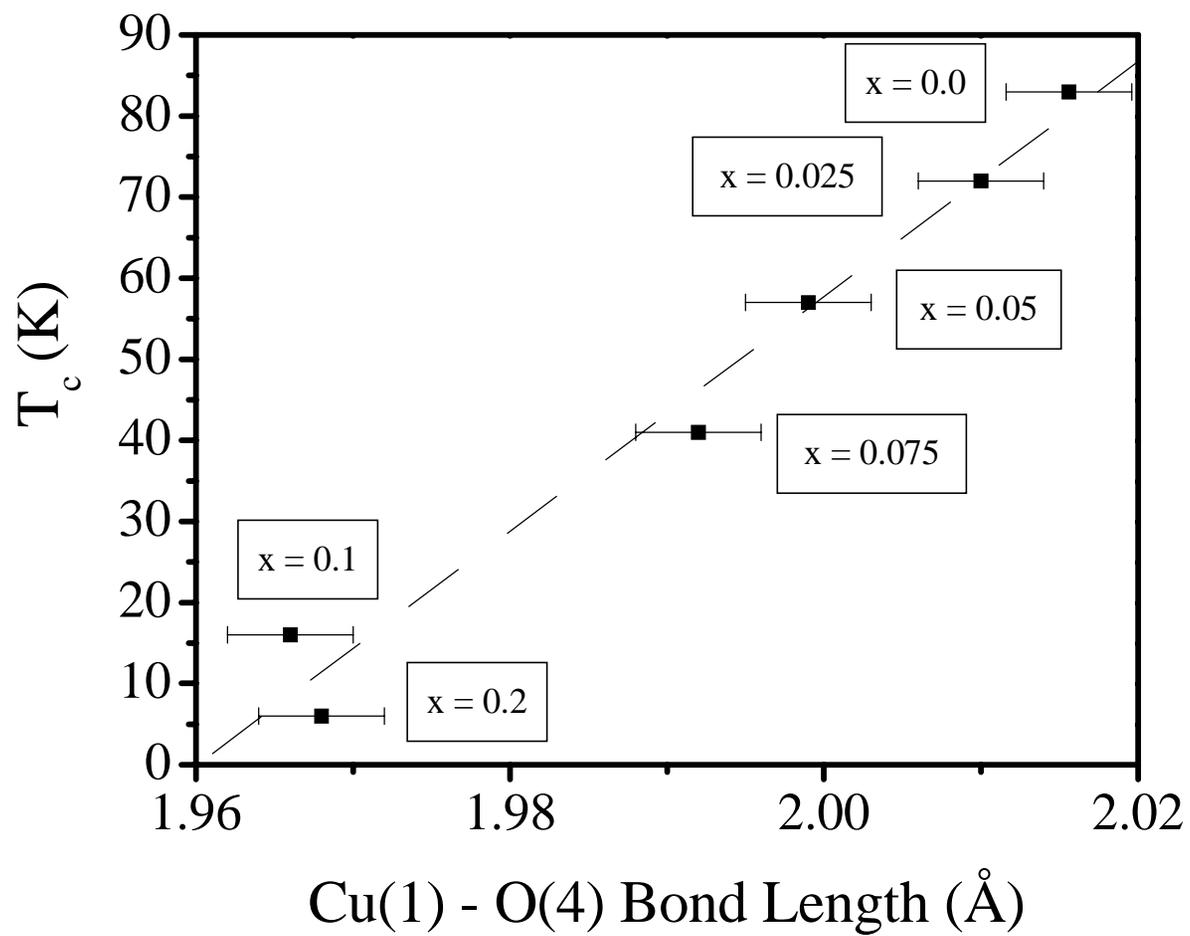

**Figure 8**

Table I (a)

Rietveld refinement results for the $(Y_{0.9}Ca_{0.1})Ba_2Cu_{4-x}Fe_xO_8$ structure with x = 0.025

| ATOM | X | Y | Z | N |
|---|---|---|---|---|
| Cu(1) | 0 | 0 | 0.2127(3) | 1.880(9) |
| Fe(1) | 0 | 0 | 0.2127(3) | 0.051(7) |
| Cu(2) | 0 | 0 | 0.0639(3) | 1.965(7) |
| Ba | ½ | ½ | 0.1344(3) | 2.0 |
| Y | ½ | ½ | 0 | 0.977(2) |
| Ca | ½ | ½ | 0 | 0.061(2) |
| O(1) | 0 | ½ | 0.2164(1) | 2.0 |
| O(2) | ½ | 0 | 0.0590(1) | 2.0 |
| O(3) | 0 | ½ | 0.0520(2) | 1.947(3) |
| O(4) | 0 | 0 | 0.1400(1) | 2.0 |

The calculated crystal parameters (in Å) are: a = 3.8461(2), b = 3.8724(2) and c = 27.256(1). Isotropic thermal parameters (in Å$^2$); 0.005 for cations, 0.010 for oxygen (see Ref. 10). The result reliability factors were: $R_{wp}$ = 10%, $R_E$ = 6.2%, $R_b$ = 11.4%, $\chi^2$ = 1.6%.

Table I (b)
Rietveld refinement results for the $(Y_{0.9}Ca_{0.1})Ba_2Cu_{4-x}Fe_xO_8$ structure with x = 0.2

| ATOM | X | Y | Z | N |
|---|---|---|---|---|
| Cu(1) | 0 | 0 | 0.2127(3) | 1.880(9) |
| Fe(1) | 0 | 0 | 0.2127(3) | 0.051(7) |
| Cu(2) | 0 | 0 | 0.0639(3) | 1.965(7) |
| Ba | ½ | ½ | 0.1345(2) | 2.0 |
| Y | ½ | ½ | 0 | 0.977(2) |
| Ca | ½ | ½ | 0 | 0.061(2) |
| O(1) | 0 | ½ | 0.2168(2) | 2.0 |
| O(2) | ½ | 0 | 0.0590(1) | 2.0 |
| O(3) | 0 | ½ | 0.0520(2) | 1.947(3) |
| O(4) | 0 | 0 | 0.1414(2) | 2.0 |
| O(5) | ½ | 0 | 0.2106(3) | 0.024(4) |

The calculated crystal parameters (in Å) are: a = 3.8472(3), b= 3.8789(2) and c = 27.205(2). Isotropic thermal parameters (in Å$^2$); 0.005 for cations, 0.010 for oxygen (see Ref. 10). The result reliability factors were: $R_{wp}$ = 5.8%, $R_E$ = 4.3%, $R_b$ = 5.9%, $\chi^2$ = 1.34%.

Rietveld refinement results for the impurity YCa123Fe structure.

| ATOM | X | Y | Z | N |
|---|---|---|---|---|
| Cu(1) | 0 | 0 | 0 | 0.979(2) |
| Fe(1) | 0 | 0 | 0 | 0.074(2) |
| Cu(2) | 0 | 0 | 0.3553(7) | 1.976(4) |
| Fe(2) | 0 | 0 | 0.3553(7) | 0.002(3) |
| Ba | ½ | ½ | 0.1840(3) | 2.0 |
| Y | ½ | ½ | ½ | 0.928(3) |
| Ca | ½ | ½ | ½ | 0.026(2) |
| O(1) | 0 | ½ | 0 | 1.775(4) |
| O(2) | ½ | 0 | 0.3686(2) | 2.0 |
| O(4) | 0 | 0 | 0.1629(2) | 2.0 |

The space group is P4/mmm and the calculated crystal parameters (in Å) are: a = b= 3.8775(9) and c = 11.6016(6). The O(2) and O(3) positions and occupation factors are equivalent in the tetragonal structure.

Table II

Mössbauer Parameters for the $Y_{0.9}Ca_{0.1}Ba_2Cu_{4-x}Fe_xO_8$ structure

| X | $IS_A$ | $\Delta Q_A$ | $\Gamma_A$ | $IS_B$ | $\Delta Q_B$ | $\Gamma_B$ | $IS_C$ | $\Delta Q_C$ | $\Gamma_C$ | $IS_D$ | $\Delta Q_D$ | $\Gamma_D$ |
|---|---|---|---|---|---|---|---|---|---|---|---|---|
| 0.025 | 0.32 | 0.796 | 0.287 | | | | | | | | | |
| 0.050 | 0.32 | 0.78 | 0.28 | | | | | | | | | |
| 0.075 | 0.32 | 0.79 | 0.29 | | | | | | | | | |
| 0.1 | 0.30 | 0.77 | 0.26 | 0.07 | 1.98 | 0.35 | 0.002 | 0.99 | 0.31 | 0.02 | 0.50 | 0.31 |
| 0.2 | 0.306 | 0.802 | 0.28 | -0.05 | 1.95 | 0.349 | -0.073 | 1.131 | 0.35 | -0.021 | 0.596 | 0.35 |

The isomer shifts is with respect to iron and all the figures are in mm/s

Mössbauer Parameters for the $YBa_2Cu_{4-x}Fe_xO_8$ structure. Ref [7]

| X | $IS_A$ | $\Delta Q_A$ | $\Gamma_A$ | $IS_B$ | $\Delta Q_B$ | $\Gamma_B$ | $IS_C$ | $\Delta Q_C$ | $\Gamma_C$ | $IS_D$ | $\Delta Q_D$ | $\Gamma_D$ |
|---|---|---|---|---|---|---|---|---|---|---|---|---|
| 0.025 | 0.32 | 0.78 | 0.28 | | | | | | | | | |
| 0.050 | 0.32 | 0.78 | 0.28 | | | | | | | | | |
| 0.075 | 0.32 | 0.79 | 0.29 | | | | | | | | | |
| 0.1 | 0.30 | 0.77 | 0.26 | 0.07 | 1.98 | 0.35 | 0.002 | 0.99 | 0.31 | 0.02 | 0.50 | 0.31 |
| 0.2 | 0.30 | 0.79 | 0.31 | 0.07 | 2.02 | 0.33 | -0.06 | 0.91 | 0.32 | -0.07 | 0.56 | 0.27 |

The isomer shifts is with respect to iron and all the figures are in mm/s